\documentstyle[aps,multicol,epsf,epsfig,psfrag,amssymb]{revtex}
\draft
\preprint{ \begin{tabular}{l}
\hbox to\hsize{September, 1999 \hfill   }\\
\hbox to\hsize{\hfill hep-th/   }\\
\end{tabular} }
\tightenlines
\begin{document}

\title{\Large\bf Glueball Mass Spectrum in KK Monopole Background}
\author{Heiyoung Shin\footnote{E-mail: hyshin@physics.utexas.edu}}
\address{Department of Physics, The University of Texas at Austin,
Austin, TX78713, USA}
\maketitle

\begin{abstract}
We consider typeIIA supergravity solution of D2-branes and D3-branes
localized within D6-branes in the near-core region of D6-branes.
With these solutions we can calculate the spectrum of the glueball
mass in QCD3 and QCD4.
The equation of motion describing the dilaton has the same eigenvalues and the
same glueball masses in QCD3 and QCD4.
Glueball mass spectrum is the same in the near core region of D6-branes of 
their M-theory counterpart is KK monopole.
We conclude that the glueball mass spectrum is the same in QCD3 and QCD4 by considering the `near-core' limit of D6-branes of which M-theory counterpart (KK monopole background)
 becomes an ALE space with an $A_{N-1}$ singularity times 7 dimensional Minkowski space $M^{(6,1)}$.
\end{abstract}

\pacs{PACS numbers:}

\begin{multicols}{2}
\narrowtext

\section{Introduction}

   Recent conjecture of relating the large $N$ behavior of certain gauge theories to {\it semi-classical} supergravity leads to a motivation for special solutions describing regions close to the cores of the brane.
In addition to this, supergravity solutions representing branes ending on branes like Hanany-Witten\cite{[1]} configuration are important.\\

In this note we discuss some explicit solutions for branes which are completely localized within other branes. The 1/4 supersymmetric solutions describe only the region close to the core of the `bigger' brane.
Supergravity solutions describing branes(D2-branes or NS 5-branes or waves)localized within D6-branes in the region close to the core of the D6-brane are already known.\cite{[2]}\\

The D2-branes localized on D6-branes are T-dual to a case of the background describing (D) strings localized on (D)5-branes and this is also related to a localized intersection of M2-branes and M5-branes. 
By U-duality and/or lifting $D=10$ solutions to $D=11$ other string theory and M-theory `near-core' localized solutions are found.\\

In section 2 we can calculate the mass spectrum of some glueballs with this supergravity solution. We found considering D6-branes in the `near-core' limit is that their M-theory counterpart (KK monopole background\cite{[3]}) becomes an ALE space with an $A_{N-1}$ singularity times a 7-dimensional Minkowski space $M^{(6,1)}$.\\

In section 3, let us consider n M2-branes along 2+1 directions in $M^{(6,1)}$.
$n$ M2-branes along 2+1 directions in $M^{(6,1)}$ are invariant under $Z_N$
identifications and the eleven-dimensional background will be given by the M2 
brane solutions with the $Z_N$ identifications in transverse space. 
In the metric obtained in this way, the $O^{++}$ glueball spectrum can be 
calculated and this calculation corresponds to the glueball mass spectrum 
in QCD3.\cite{[4]}
This M2-brane solution is related to ten dimensions to obtain a typeIIA solution which can be interpreted as the near-core region of a configuration of D2-branes localized within a collection of D6-branes.\\

In section 4, we also suppose $n$ M3-branes along 3+1 directions in $M^{(6,1)}$. This M3 branes along 3+1 directions in $M^{(6,1)}$ are invariant under $Z_N$ identifications and eleven dimensional background will be given by the M3 solutions with the $Z_N$ identification in transverse space.
The glueball mass spectrum in this supergravity solution corresponds to glueball mass in QCD4.\\

The dilaton equations of motion are the same in the 3 near-core limits of their M-theory counterpart is KK monopole. In conclusion, the glueball mass spectrum in QCD3 and QCD4 has the same results in this near-core limits and this mass spectrum is unique for this background.

\section{D2-branes localized on D6-branes}

  In the near core region of typeIIA solution for a collection of $N$ D6-branes upon dimension reduction of the $A_{N-1}$ space\cite{[5]}, the metric is\\
 \begin{eqnarray}
   ds_{11}^2&=&dx_\|^2+d\rho^2+\rho^2(d\bar{\theta}^2+sin^2\bar{\theta}d\bar{\varphi}^2+cos^2\bar{\theta}d\bar{\phi}^2)\\   
      dx_\|^2&=&-dt^2+dx_1^2+\cdots+dx_6^2\\
        \rho^2&=&x_7^2+\cdots+x_{10}^2\\
            0&\le&\bar{\theta}\le\pi/2, and\\
            0&\le&\bar{\varphi},\bar{\phi}\le2\pi\\
\end{eqnarray}

with the $Z_N$ identification of $(\bar{\varphi},\bar{\phi})\sim(\bar{\varphi},\bar{\phi})+(\frac{2\pi}{N},\frac{2\pi}{N})$.
If we define new variables,

\begin{eqnarray}
  Y=\frac{\rho^2}{2Nl_p^3},
   \theta=2\bar{\theta},
   \varphi=\bar{\varphi}-\bar{\phi},
   \phi=N\bar{\phi},                                
\end{eqnarray}

then the metric is
\begin{eqnarray} 
ds_{11}^2&=&dx_\|^2+\frac{Nl_p^3}{2Y}dY^2+\frac{Nl_p^3Y}{2}(d\theta^2+sin^2\theta d\varphi^2)\\
     &+&\frac{2Yl_p^3}{N}(d\phi+\frac{N}{2}(cos\theta-1)d\varphi)^2\nonumber   
\end{eqnarray}
, where $\phi$ has the period of $2\pi$.\\

After the reduction to ten dimensions by using Killing vector along the $\phi$ direction, we obtain ten dimesional typeIIA string metric, dilaton, and gauge field.
\begin{eqnarray}
ds_{10}^2&=&\alpha'\left[\frac{(2\pi)^2}{g_{YM}}\sqrt{\frac{2Y}{N}}dx_\|^2+\frac{g_{YM}}{(2\pi)^2}\sqrt{\frac{N}{2Y}}dY^2\right] \nonumber\\
         &+&\alpha'\left[\frac{g_{YM}}{(2\pi)^2}\sqrt{\frac{N}{2}}Y^{\frac{3}{2}}d\Omega_2^2\right]\\
         e^{\phi}&=&\frac{g_{YM}^2}{2\pi}\left(\frac{2Y}{g_{YM}^2N}\right)^{\frac{3}{4}}\\
         A_{\mu}dx^{\mu}&=&\frac{N}{2}(cos\theta-1)d\varphi
\end{eqnarray}

, where $d\Omega^2=d\theta^2+sin^2{\theta}d\varphi^2$ and the relation to eleven dimensional metric is
\begin{eqnarray}
   ds_{11}^2&=&e^{\frac{4}{3}\phi}(dx_{11}+A_{\mu}dx^{\mu})^2+e^{-\frac{2}{3}\phi}ds_{10}^2.
\end{eqnarray}
Equation (9) is the typeIIA solution for D6-branes in the decoupling limit\cite{[6]}
\begin{eqnarray}
  Y=\frac{|x|}{\alpha'}&=&fixed,\\
  g_{YM}^2=(2\pi)^4l_p^3&=&(2\pi)^4g_s\alpha'^{\frac{3}{2}}=fixed,\\
         &\alpha'&\rightarrow 0.     
\end{eqnarray}

In this metric, we can obtain the equation of motion for the dilaton,
\begin{eqnarray}
\partial_\mu[e^{-2\phi}\sqrt{g}g^{\mu\nu}\partial_\nu\Phi]&=&0
\end{eqnarray}

If we assume $\Phi=f(Y)e^{(ik \cdot x)}$  and plug this into equation(16) and by the metric, we obtain the following equation for $f$.
\begin{eqnarray}
  \partial_Y[Y^2\partial_Yf]-\frac{g_{YM}^2Nk^2Y}{2(2\pi)^4}f&=&0   
\end{eqnarray}
where glueball mass $M^2$ is equal to $-k^2$.
Since $Y=0$ is regular and $Y=\infty$ is irregular singular point, we assume $f$ in the following form.
\begin{eqnarray}
   f &=& \sum_{\lambda} a_{\lambda} Y^{\lambda}       
\end{eqnarray}

  After plugging this into the equation (17), we obtain the following recurrence relation.
\begin{eqnarray}
   a_\lambda&=&\frac{g_{YM}^2Nk^2a_{\lambda-1}}{2(2\pi)^4\lambda(\lambda+1)}  
\end{eqnarray}

  Since normalization is arbitrary, we set $a_0=1$ and the first few coefficients are 
\begin{eqnarray}
a_1&=&\frac{g_{YM}^2Nk^2}{4(2\pi)^4}\\
a_2&=&{\left(\frac{g_{YM}^2Nk^2}{2(2\pi)^4}\right)}^2\frac{1}{12}\\
a_3&=&{\left(\frac{g_{YM}^2Nk^2}{2(2\pi)^4}\right)}^3\frac{1}{144}   
\end{eqnarray}

Since $f$ has the unique solution for a certain value of $k^2$, this is an eigenvalue problem. The $O^{++}$ glueball mass $M^2$ is $-k^2$.

\section{D2-branes within D6-branes}

  Let us consider a configuration of M2-branes stretched along $x_1,x_2$ directions.\cite{[7]} In this situation we shall put all of them at the origin in the transverse space. In this case, the metric is 
\begin{eqnarray}
ds_{11}^2&=&f_2^{-\frac{2}{3}}(-dt^2+dx_1^2+dx_2^2)\\
         &+&f_2^{\frac{1}{3}}[dx_3^2+dx_4^2+dx_5^2+dx_6^2 \nonumber\\
         &+&\rho^2(d\bar{\theta}^2+sin^2\bar{\theta}d\bar{\varphi}^2+cos^2\bar{\theta}d\bar{\phi}^2)] \nonumber        
\end{eqnarray}

, where
 \begin{eqnarray}
f_2&=&1+\frac{2^5\pi^2n_2l_p^6}{\hat{r}^6}\\
  \hat{r}^2&=&(x_3^2+x_4^2+x_5^2+x_6^2+\rho^2)\\
  \rho^2&=&x_7^2+\cdots+x_{10}^2
\end{eqnarray}

 and the 3-rank tensor has the standard form $C_{012}=f_2^{-1}$.
If we use the new variables of equation (7) , then the above metric(23) is
\begin{eqnarray}
ds_{11}^2&=&f_2^{-\frac{2}{3}}(-dt^2+dx_1^2+dx_2^2)\\
        &+&f_2^{\frac{1}{3}}[dx_3^2+dx_4^2+dx_5^2+dx_6^2+\frac{Nl_p^3}{2Y}dY^2 \nonumber\\
         &+&\frac{Nl_p^3Y}{2}(d\theta^2+sin^2\theta d\varphi^2) \nonumber\\
	 &+&\frac{2l_p^3Y}{N}(d\phi+\frac{N}{2}(cos\theta-1)d\varphi)^2] \nonumber
\end{eqnarray}

, where
\begin{eqnarray}
   f_2&=&1+\frac{2^5\pi^2n^2l_p^6}{(x_3^2+x_4^2+x_5^2+x_6^2+2Nl_p^3Y)^3}
\end{eqnarray}

After dimension reduction to $D=10$ along the $\phi$ direction, we can obtain the following metric
\begin{eqnarray}
ds_{10}^2&=&\alpha'[f_2^{-\frac{1}{2}}h_6^{-\frac{1}{2}}(-dt^2+dx_1^2+dx_2^2)\\
         &+&f_2^{\frac{1}{2}}h_6^{-\frac{1}{2}}(dx_3^2+dx_4^2+dx_5^2+dx_6^2) \nonumber\\
         &+&f_2^{\frac{1}{2}}h_6^{\frac{1}{2}}(dY^2+Y^2d\Omega^2)] \nonumber\\
   e^\phi&=&\frac{g_{YM}^2}{(2\pi)^4}f_2^{\frac{1}{4}}h_6^{-\frac{3}{4}}\\
   A_{\mu}dx^{\mu}&=&\frac{N}{2}(cos\theta-1)d\phi\\
   h_6&=&\frac{g_{YM}^2N}{2(2\pi)^4Y}
\end{eqnarray}

where $x_3,x_4,x_5,x_6$ are the coordinates along the D6-branes where the QCD4 lives,while Y is the radial coordinate of the space. In this metric, we can obtain the equation  of motion for the dialton,
\begin{eqnarray}
   \partial_\mu[e^{-2\phi}\sqrt{g}g^{\mu\nu}\partial_\nu\Phi]&=&0.  
\end{eqnarray}

As a result we can get the same equation(17) for the dilaton and the equation is

\begin{eqnarray}
\partial_Y(Y^2\partial_Y f)-\frac{g_{YM}^2Nk^2Y}{2(2\pi)^4}f&=&0.   
\end{eqnarray}

Assuming $\Phi=f(Y)e^{ik \cdot x}$, we get the same eigenvalue of $k^2$ for the above equation and $O^{++}$ glueball mass $M^2$ is equal to $-k^2$.

\section{D3-branes within D6-branes}

Let us consider a configuration of M3-branes stretched along $x_1,x_2,x_3$ directions.\cite{[8]},\cite{[9]} The metric is
\begin{eqnarray}
ds_{11}^2&=&f_3^{-\frac{5}{9}}(-dt^2+dx_1^2+dx_2^2+dx_3^2)\\
         &+&f_3^{\frac{4}{9}}[dx_4^2+dx_5^2+dx_6^2  \nonumber\\
	 &+&\rho^2(d\bar{\theta}^2+sin^2\bar{\theta}d\bar{\varphi}^2+cos^2\bar{\theta}d\bar{\phi}^2)] \nonumber  
\end{eqnarray}

where
\begin{eqnarray}
f_3&=&\frac{Q^3}{\hat{r}^3}\\
  \hat{r}^2&=&x_4^2+x_5^2+x_6^2+\rho^2\\
  \rho^2&=&x_7^2+x_8^2+x_9^2+x_{10}^2
\end{eqnarray}

Using the new variables of equation (7), the above metric is
\begin{eqnarray}
ds_{11}^2&=&f_3^{-\frac{5}{9}}(-dt^2+dx_1^2+dx_2^2+dx_3^2)\\
         &+&f_3^{\frac{4}{9}}[dx_4^2+dx_5^2+dx_6^2+\frac{Nl_p^3}{2Y}dY^2  \nonumber\\
         &+&\frac{Nl_p^3Y}{2}(d\theta^2+sin^2\theta d\varphi^2)  \nonumber\\
	 &+&\frac{2l_p^3Y}{N}(d\phi+\frac{N}{2}(cos\theta-1)d\varphi)^2]  \nonumber 
\end{eqnarray}

,where
\begin{eqnarray}
f_3&=&\frac{Q^3}{(x_4^2+x_5^2+x_6^2+2Nl_p^3Y)^3}
\end{eqnarray}

After the dimension reduction to $D=10$ along the $\phi$ direction, we obtain the following typeIIA supergravity solution for D3-branes localized within D6-branes.
\begin{eqnarray}
ds_{10}^2&=&\alpha'[f_3^{-\frac{1}{3}}h_6^{-\frac{1}{2}}(-dt^2+dx_1^2+dx_2^2+dx_3^2)\\
         &+&f_3^{\frac{2}{3}}h_6^{-\frac{1}{2}}(dx_4^2+dx_5^2+dx_6^2) \nonumber\\
         &+&f_3^{\frac{2}{3}}h_6^{\frac{1}{2}}(dY^2+Y^2d\Omega^2)] \nonumber\\
   e^\phi&=&\frac{g_{YM}^2}{(2\pi)^4}f_3^{\frac{1}{3}}h_6^{-\frac{3}{4}}\\
  A_{\mu}dx^\mu&=&\frac{N}{2}(cos\theta-1)d\varphi\\
  h_6&=&\frac{g_{YM}^2N}{2(2\pi)^4Y}
\end{eqnarray}

,where $x_4,x_5,x_6$ are the coordinates along D6-brane where QCD3 lives while Y is the radial coordinate of the space.
If we follow the same procedure for the D2+D6 case, we can conclude that the equation of motion  for the dilaton is the same form as equation (17).\\

\section{Conclusion}

  We have found that how glueball mass spectrum can be derived from the Maldacena's conjecture in the near-core limit. The glueball mass spectrum is the same for QCD3 and QCD4 in the KK monopole background.

In my next research, I will do calculations  of the glueball mass spectrum as an eigenvalue problem of the equation of motion  for dilaton in this work.
The result can be compared with the lattice calculation.\cite{[10]},\cite{[11]}

It would be very important to understand whether the coincidence of glueball mass spectrum for QCD3 and QCD4 is only for KK monopole background or whether this  fact is for the same background.

\acknowledgments
I thank Philip Candelas for several discussions, based on which this work has been done.

\end{multicols}

\begin{references}
\bibitem{[1]}  J.Maldacena {\it The Large  N Limit of Superconformal Field Theories and Supergravity}, hep-th/9711200.


\bibitem{[2]}  N.Itzhaki, A.A.Tseylin and S.Yankielowicz,{\it Supergravity Solution for Branes Localized Within Branes}, hep-th/9803103.


\bibitem{[3]}  P.K.Townsend,{\it Eleven-Dimensional Supermembrane Revisited}, 
Phys .Lett .B {\bf 350} (1995) p.184, hep-th/9501068.


\bibitem{[4]}  Csaba Csaki, Jorge Russo, Konstandinos Sfetsos and John Terning, {\it Supergravity Models for 3+1 Dimensional QCD}, hep-th/9902067.


\bibitem{[5]}  N.Itzhaki, J.Maldacena, J.Sonnenschein and S.Yankielowicz, {\it Supergravity and The Large N  Limit of Theories With Sixteen Supercharges}, hep-th/9802042.


\bibitem{[6]}  J.Maldacena,{\it Branes Probing Black Holes}, hep-th/9709099.


\bibitem{[7]}  M.J.Duff and  K.S.Stelle, {\it Multimembrane Solutions of D = 11 Supergravity}, Phys. Lett. B {\bf 253} (1991) p.113.


\bibitem{[8]}  A.A.Tseylin, {\it Composite BPS Configurations of D-branes in 10 and 11 Dimensions}, Class. Quant. Grav. {\bf 14} (1997) p.2085, hep-th/9702163.


\bibitem{[9]}  A.A.Tseylin, {\it Harmonic Superposition of M-branes}, Nucl.Phys.B {\bf 475} (1996) p.149, hep-th/9604035.


\bibitem{[10]}  C.Morningstar and M.Peardon, Phys. Rev. D {\bf 56} p.4043 (1997).


\bibitem{[11]}  M.Peardon, Nucl. Phys. B {\bf 63} 22 (1998); C.Morningstar and M.Peardon, hep-lat/9901004.
\end{references}
\end{document}